\documentclass[aps,pre,twocolumn,superscriptaddress,showpacs]{revtex4}
\usepackage{amsfonts,amssymb,amsmath,latexsym,epsfig} 
\begin{document}

[AIP Conference Proceedings {\bf 776}, 201 (2005)]\\

\title{Weighted networks are more synchronizable: how and why}

\author{Adilson E. Motter}
\email{motter@mpipks-dresden.mpg.de}
\affiliation{Max Planck Institute for the Physics of Complex Systems,
N\"othnitzer Str. 38, 01187 Dresden, Germany}

\author{Changsong Zhou}
\email{cszhou@agnld.uni-potsdam.de}
\affiliation{Institute of Physics, University of Potsdam
PF 601553, 14415 Potsdam,  Germany}

\author{J\"{u}rgen Kurths}
\affiliation{Institute of Physics, University of Potsdam
PF 601553, 14415 Potsdam,  Germany}

\date{\today}

\begin{abstract} Most real-world networks display not only a heterogeneous
distribution of degrees, but also a heterogeneous distribution of weights in the
strengths of the connections.  Each of these heterogeneities alone has been shown to
suppress synchronization in random networks of dynamical systems.  Here we
review our recent findings that complete synchronization is significantly
enhanced and becomes independent of both distributions when the distribution of
weights is suitably combined with the distribution of degrees.  We also present
new results addressing the optimality of our findings and extending our analysis to
phase synchronization in networks of non-identical dynamical units.

\end{abstract}



\maketitle

\section{Introduction}

The recent explosion in the study of complex networks of dynamical systems~\cite{rev}
has provided compeling evidence that the synchronization of coupled oscillators
is drastically influenced by the structure of the underlying network of
couplings~\cite{watts:book,sync,lgh,BP:2002,NMLH:2003,GDLP:2000,Wu:2003,heter,ROH:2004}.
Previous works on synchronization have focused mainly on the influence of the
topology of the connections by assuming that the coupling strength is uniform.
These works have shown that the ability to synchronize is generally enhanced in
both small-world networks (SWNs) \cite{sw} and random scale-free networks (SFNs) \cite{sf}
as compared to regular networks \cite{com1}, indicating that synchronizability is
strongly related to the average distance between oscillators.  More recently, it
has been shown that synchronizability also depends critically on the
heterogeneity of the degree distribution and that this can lead to
counter-intuitive behaviors \cite{NMLH:2003}.  In particular, random networks with
strong heterogeneity in the degree distribution, such as SFNs, are more
difficult to synchronize than random homogeneous networks \cite{NMLH:2003},
despite the fact that heterogeneity reduces the average distance between nodes \cite{spath}.

However, synchronization, as many other dynamical processes, is influenced not only by
the topology, but also by the strength of the connections.  Very recently, it
has been shown that synchronization is suppressed in networks with heterogenous
distribution of weights in the connection strengths \cite{heter}, even when the
distribution of degrees is homogeneous.  This is an important result because
most complex networks where synchronization is relevant are indeed weighted and
display a highly heterogeneous distribution of both degrees {\it and}
weights~\cite{BBPV:2004,MAB:2004,w}.  Examples include brain networks, both at the neuronal
and cortical level~\cite{brain}, and airport networks~\cite{BBPV:2004}, which
underlie the synchronization of epidemic outbreaks in different cities~\cite{city}.
The identification and study of complex weighted networks with improved synchronization
properties is thus of great interest.

In this paper, we review this fundamental problem in the context of complete
synchronization of identical oscillators~\cite{MCJ1,MCJ2}. We introduce a weighted
coupling scheme  and we show that, for a given network topology, the
synchronizability is maximum when the network of couplings is weighted
and directed. For large sufficiently random networks, the maximum
synchronizability is primarily determined by the mean degree of the network and
does not depend on the degree distribution and system size.  This contrasts with
the case of unweighted (and undirected) coupling, where the synchronizability is
strongly suppressed as the degree heterogeneity or number of oscillators is increased.
We also show that the total cost involved in the weighted couplings is
significantly reduced as compared to the case of unweighted coupling and 
appears to be minimum when the synchronizability is maximum.
Furthermore, we present new results that extend our findings to the synchronization
of phase oscillators and provide evidence for the optimality of our weighted model.

The paper is organized as follows.  The problem of complete synchronization is
formulated in Sec.~\ref{s2} and is analyzed in Sec.~\ref{s3}.  The problem of
phase synchronization is considered in Sec.~\ref{s4}.  In Sec.~\ref{s5}, we
discuss the distribution of weights that would maximize synchronizability for a
given network topology.  The conclusions are presented in the last section.

\section{Complete synchronization}
\label{s2}

We introduce a weighted model of linearly coupled identical oscillators and
we present a condition for the linear stability of the completely synchronized
states in terms of the eigenvalues of the coupling matrix.

The dynamics of a weighted network of $N$ identical oscillators is described by:
\begin{eqnarray}
\dot{\bf x}_i&=&{\bf F}({\bf x}_i)+\sigma\sum_{j=1}^{N}A_{ij} [{\bf H}({\bf x}_j)-{\bf H}({\bf x}_i)],\label{eq1}\\
         &=&{\bf F}({\bf x}_i)-\sigma\sum_{j=1}^{N}G_{ij} {\bf H}({\bf x}_j), \;\;\; i=1,\ldots ,N, 
\label{eq1}
\end{eqnarray}
where ${\bf F}={\bf F}({\bf x})$ governs the dynamics of each individual oscillator, ${\bf H}={\bf H}({\bf x})$ is
the output function,  and $\sigma$ is the overall coupling strength. 
Matrix $A=(A_{ij})$ is the adjacency matrix of the underlying network of couplings,
where $A_{ij} = w_{ij}$ if there is a link of strength $w_{ij}>0$ from node $j$ to node $i$, and $0$ otherwise.
Matrix $G=(G_{ij})$, defined as $G_{ij}= -A_{ij} +\delta_{ij}\sum_{j=1}^{N}A_{ij}$, is the coupling matrix.
The rows of matrix $G$ 
have zero sum and this ensures that the completely synchronized state 
$\{ {\bf x}_i(t)={\bf s}(t), \forall i \; | \; d{\bf s}/dt={\bf F}({\bf s}) \}$
is a solution of Eq.~(\ref{eq1}).

In the case of symmetrically coupled oscillators with uniform coupling strength,
the network of couplings is unweighted and undirected, and
$G$ is the usual (symmetric) Laplacian matrix $L=(L_{ij})$:  the diagonal
entries are $L_{ii} = k_i$, where $k_i$ is the degree of node $i$, and the
off-diagonal entries are $L_{ij}=-1$ if nodes $i$ and $j$ are connected and
$L_{ij}=0$ otherwise.  For $G_{ij}=L_{ij}$, heterogeneity in
the degree distribution suppresses synchronization in important
classes of networks \cite{NMLH:2003}.  The synchronizability may be easily enhanced if we modify
the topology of the network of couplings.  Here, however, we address the problem
of enhancement of synchronizability for a {\it given} network topology.

\begin{figure}[pt]
\begin{center}
\epsfig{figure=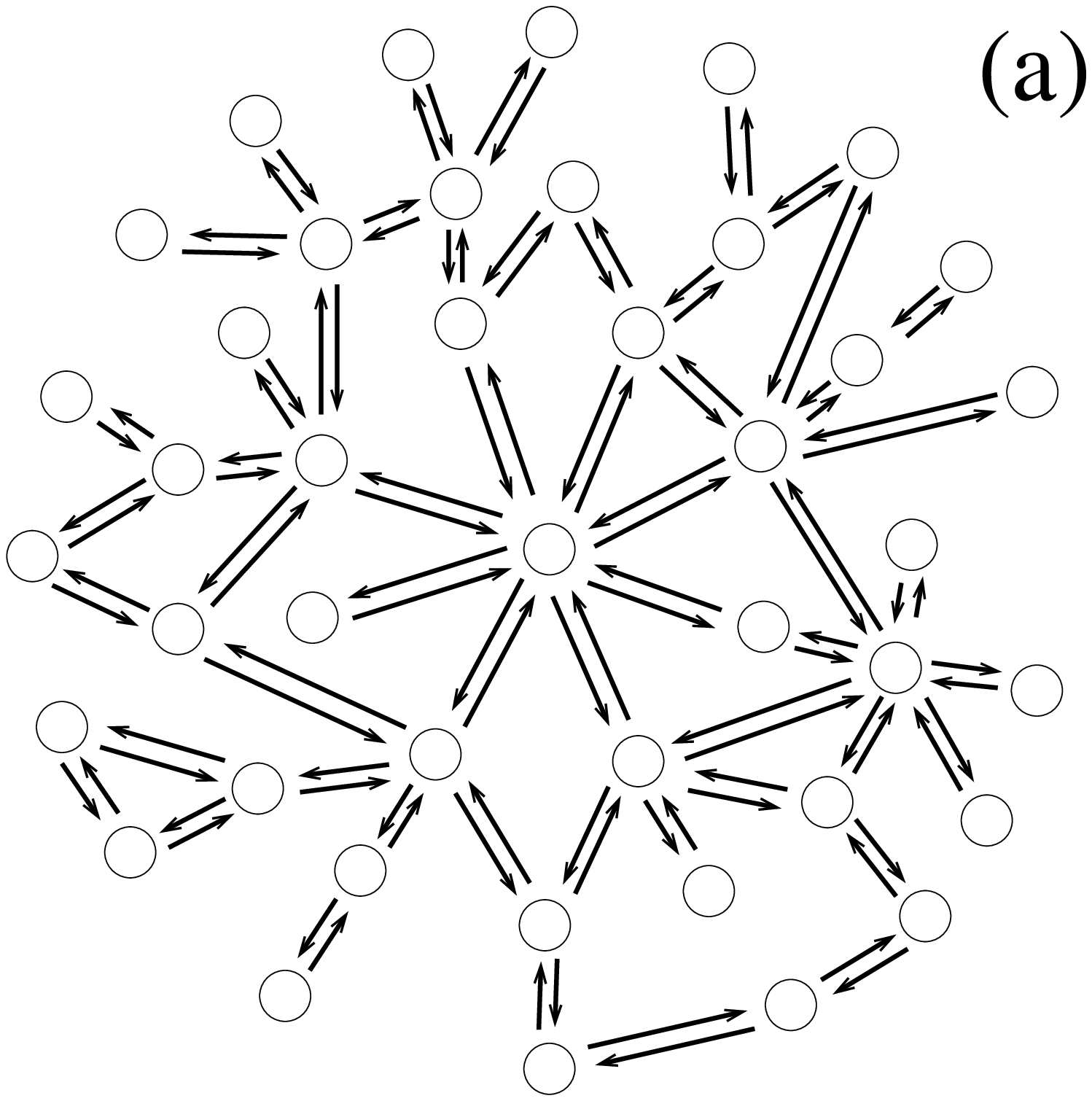,width=6.0cm}
\epsfig{figure=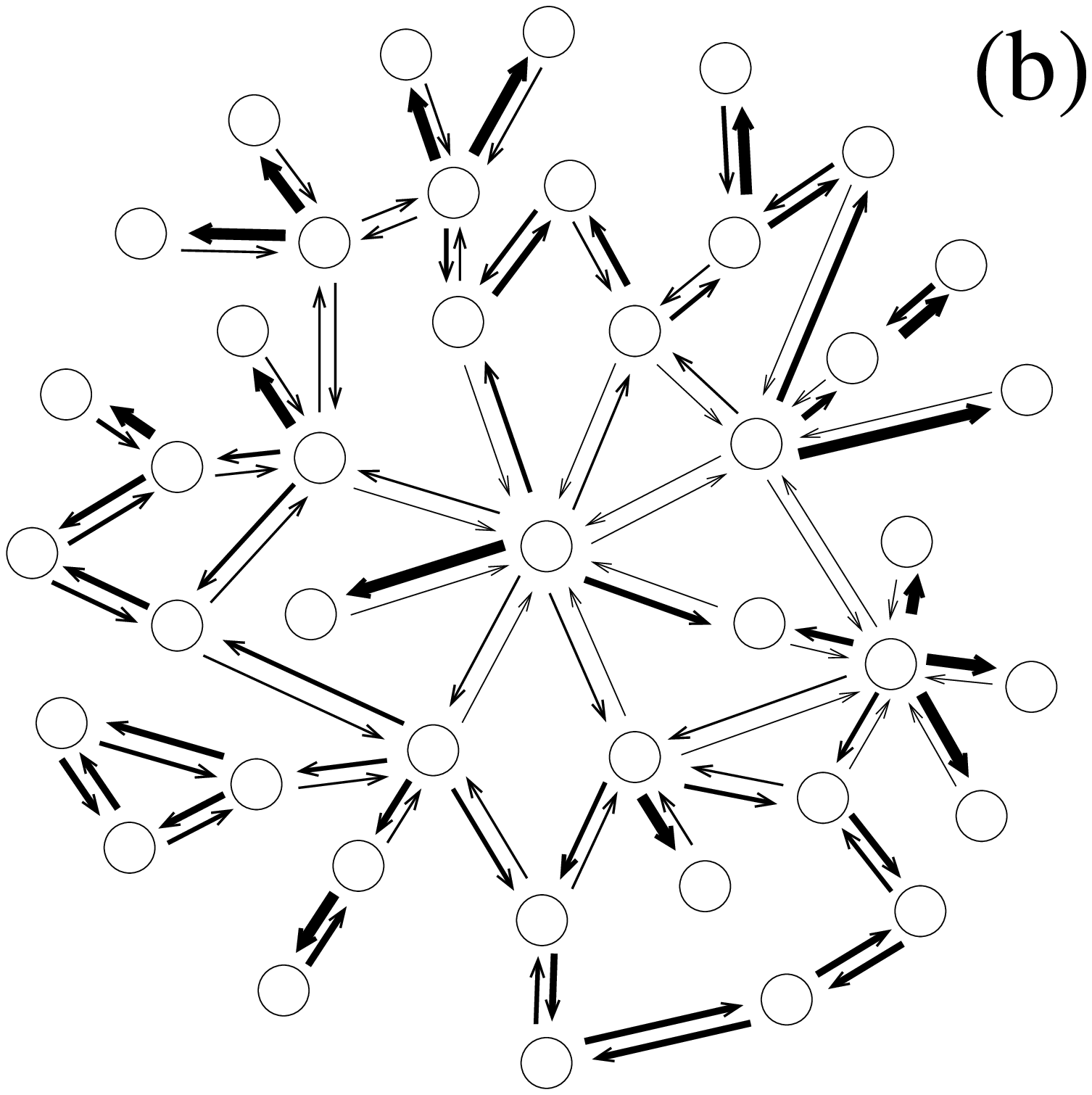,width=6.0cm}
\caption{Illustration of model (\ref{eq2}) for (a) $\beta=0$ (unweighted coupling) and (b) $\beta>0$
(weighted coupling).}
\label{figI}
\end{center}
\end{figure}

In order to enhance the synchronizability of networks with heterogeneous degree
distribution, we propose
to scale the coupling strength by a function of the degree of the nodes. 
For specificity, we take
\begin{equation}
G_{ij}=L_{ij}/k_i^{\beta},
\label{eq2}
\end{equation}
where $\beta$ is a tunable parameter.  We say that the network or coupling is
weighted when $\beta\neq 0$ and unweighted when $\beta=0$. Networks with $\beta=0$ 
and $\beta>0$ are depicted in Figs.~\ref{figI}(a) and \ref{figI}(b), respectively.

The underlying network associated with the Laplacian matrix $L$ is undirected and unweighted,
but for $\beta\neq 0$, the network of couplings becomes not only weighted but
also directed because the resulting matrix $G$ is in general asymmetric. 
This is a special kind of directed network where the number of {\it
in}-links is equal to the number of {\it out}-links in each node, and the
directions are encoded in the strengths of in- and out-links [Fig.~\ref{figI}(b)].  In spite of 
the possible asymmetry of matrix $G$,
all the eigenvalues of matrix $G$ are nonnegative reals and can be ordered as
$0=\lambda_1\le \lambda_2\cdots \le\lambda_N$, as shown below.

In matrix notation, Eq.~(\ref{eq2}) can be written as $ G=D^{-\beta}L$, where
$D=$ diag$\{k_1,k_2,\ldots k_N\}$ is the diagonal matrix of degrees. 
(We recall that the degree $k_i$ is the number of
oscillators coupled to oscillator $i$.) From the
identity $\det (D^{-\beta}L-\lambda I)=\det (D^{-\beta/2}LD^{-\beta/2}-\lambda
I)$, valid for any $\lambda$, we have that the spectrum of eigenvalues of matrix
$G$ is equal to the spectrum of a symmetric matrix defined as
$H=D^{-\beta/2}LD^{-\beta/2}$.  As a result, all the eigenvalues of matrix $G$
are real, as anticipated above.  Moreover, because $H$ is positive semidefinite, all the eigenvalues
are nonnegative and, because the rows of $G$ have zero sum, the smallest
eigenvalue $\lambda_1$ is always zero.  If the network is
connected, $\lambda_2>0$ for any finite $\beta$. Naturally,
the study of complete synchronization of the whole network only makes sense if the
network is connected.  
For $\beta=1$, matrix $H$ is the normalized Laplacian matrix  \cite{chung:book}. 
In this case, if $N \ge 2$ and the network is
connected, then $0<\lambda_2\le N/(N-1)$ and $ N/(N-1)\le\lambda_N\le 2$.
For spectral properties of
unweighted complex networks,
see Refs.~\cite{chung:book,other_ref,CLV:2003,DGMS:2003}.

The variational equations governing the linear stability of a synchronized state
$\{ {\bf x}_i(t)={\bf s}(t), \forall i\}$ of the system in Eqs.~(\ref{eq1}) and (\ref{eq2})
can be diagonalized into $N$ blocks of the form
\begin{equation}
\frac{d\eta}{dt}=\left[ D{\bf F}({\bf s}) - \alpha D{\bf H}({\bf s})\right]\eta,
\label{eq6}
\end{equation}
where $D$ denotes the Jacobian matrix, $\alpha=\sigma\lambda_i$, and $\lambda_i$
are the eigenvalues of the coupling matrix $G$.  The largest Lyapunov exponent
$\Lambda(\alpha)$ of this equation can be regarded as a master stability
function, which determines the linear stability of the synchronized state
\cite{msf}:  the synchronized state is stable if
$\Lambda(\sigma\lambda_i)<0$ for $i=2,\ldots N$.  
(The eigenvalue $\lambda_1$ corresponds to a mode parallel to the synchronization manifold.)

\begin{figure}[pt]
\begin{center}
\epsfig{figure=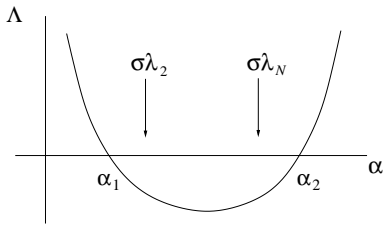,width=5.0cm}
\caption{Master stability function: illustration of the condition
$\alpha_1 < \sigma\lambda_2 \le \ldots \le \sigma\lambda_N <\alpha_2$.}
\label{figII}
\end{center}
\end{figure}

For many oscillatory dynamical systems \cite{BP:2002,msf}, the master
stability function $\Lambda(\alpha)$ is negative in a single, finite interval
$(\alpha_1,\alpha_2)$. Therefore,
the network is synchronizable for some
$\sigma$ when $\alpha_1 < \sigma\lambda_2 \le \ldots \le \sigma\lambda_N <\alpha_2$,
as depicted in Fig.~\ref{figII}.
This is equivalent to the condition
\begin{equation}
R\equiv\lambda_N/\lambda_2<\alpha_2/\alpha_1,
\label{eq7}
\end{equation}
where $\alpha_2/\alpha_1$ depends only on the dynamics (${\bf F}$, ${\bf H}$, and
${\bf s}$), while the eigenratio $R$ depends only on the coupling matrix $G$.  The
problem of synchronization is then reduced to the analysis of eigenvalues of the
coupling matrix \cite{BP:2002}:  the smaller the eigenratio $R$, the larger the
synchronizability of the network and vice versa. Alternatively, the eigenratio $R$ can
also be regarded as a measure of the {\it speed} to synchronize, where the speed
is defined by the smallest transverse Lyapunov exponent.

\section{Enhancement of synchronizability in weighted networks}
\label{s3}

We study the system  in Eqs.~(\ref{eq1}) and (\ref{eq2})
and we argue that the synchronizability is maximum ($R$ is minimum)
and the coupling cost is minimum for $\beta=1$.

\subsection{Mean Field Approximation}

A mean field approximation provides insight into the effects of degree
heterogeneity and the dependence of $R$ on $\beta$.

The dynamical equations (\ref{eq1}) can be rewritten as:
\begin{equation}
\frac{d{\bf x}_i}{dt}={\bf F}({\bf x}_i)+\sigma k_i^{1-\beta} [\bar {\bf H}_i-{\bf H}({\bf x}_i)],
\label{eq13}
\end{equation}
where 
\begin{equation} 
\bar {\bf H}_i=\frac{1}{k_i} \sum_j A_{ij} {\bf H}({\bf x}_j)
\label{eq14}
\end{equation}
is the local mean field from all the nearest neighbors of oscillator $i$.
If the network is sufficiently random and the system is close to the
synchronized state ${\bf s}$, we may assume that $\bar {\bf H}_i\approx {\bf H}({\bf s})$ and we may
approximate Eq.~(\ref{eq13}) as
\begin{equation}
\frac{d{\bf x}_i}{dt}={\bf F}({\bf x}_i)+\sigma k_i^{1-\beta}[{\bf H}({\bf s})-{\bf H}({\bf x}_i)],
\label{eq15}
\end{equation}
indicating that all the oscillators are decoupled and forced by 
a common  oscillator with output ${\bf H}({\bf s})$. 

From a variational equation analogous to Eq.~(\ref{eq6}), we have that all
oscillators in Eq.~(\ref{eq15}) will be synchronized by the common forcing when
\begin{equation} 
\alpha_1<\sigma k_i^{1-\beta}<\alpha_2 \;\;\; \forall i.
\label{eq16}
\end{equation}
For $\beta\ne 1$, it is enough to have a single node with degree very different
from the others for this condition not to be satisfied for any $\sigma$.  In
this case, the complete synchronization becomes impossible because the
corresponding oscillator cannot be synchronized. This explains the results
of Ref.~\cite{NMLH:2003} on the suppression of synchronizability due to
heterogeneity in unweighted networks.

Within the mean field approximation, the eigenratio is
$R=(k_{\max}/k_{\min})^{1-\beta}$ for $\beta\leq 1$ and
$R=(k_{\min}/k_{\max})^{1-\beta}$ for $\beta > 1$, where $k_{\min}=\min_i \{ k_i \}$
and $k_{\max}=\max_i \{ k_i \}$, and reaches its minimum at $\beta=1$.
%
This suggests that the maximum
synchronizability in weighted networks is achieved when $\beta=1$.  As shown in
Ref.~\cite{MCJ2}, the same indication is provided by the analysis of a diffusion
process related to the spread of information in the network.  We thus expect
the (un-approximated) system in Eq.~(\ref{eq1}) to exhibit maximum synchronizability
at $\beta=1$.

We now test this prediction numerically on random SFNs \cite{NSW:2001}.
The networks are generated as follows.  Each node is assigned
to have a number $k_i \ge k_{\min}$ of ``half-links'' according to the
probability distribution $P(k)\sim k^{-\gamma}$, where $\gamma$ is a scaling
exponent and $k_{\min}$ is a constant integer.  The network is generated by
randomly connecting these half-links to form links, prohibiting self- and
repeated links.  We discard the networks that are not connected.
In the limit $\gamma=\infty$, all nodes have the same degree $k= k_{\min}$.

\begin{figure}[pt]
\begin{center}
\epsfig{figure=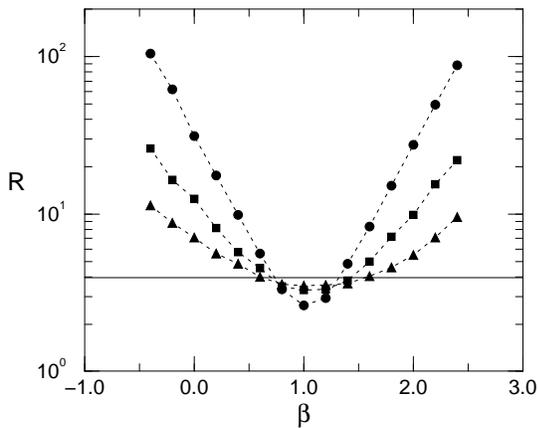,width=7.0cm}
\caption{
Eigenratio $R$ as a function of $\beta$ for random SFNs with $\gamma=3$
($\bullet$), $\gamma=5$ (${\scriptscriptstyle \blacksquare}$),
$\gamma=7$ (${\scriptstyle \blacktriangle}$),  and $\gamma=\infty$ (solid line).
Each curve corresponds to an average over $50$ realizations of the networks for
$k_{\min}=10$ and $N=1024$.
}
\label{fig1}
\end{center}
\end{figure}

Our numerical computations confirm that the eigenratio $R$ has a well
defined minimum at $\beta=1$, as shown in Fig.~\ref{fig1}
for various $\gamma$.  The only exception is the class
of homogeneous networks, where all the nodes have the same degree.  When the
network is homogeneous, the weights $k_i^{-\beta}$ can be factored out
and the eigenratio $R$ is independent of $\beta$ (Fig.~\ref{fig1}, solid line).
Random homogeneous networks correspond to random SFNs with $\gamma=\infty$.
In SFNs, the heterogeneity increases as the scaling exponent $\gamma$ is
reduced.  As shown in Fig.~\ref{fig1}, the minimum of the
eigenratio $R$ becomes more pronounced as the heterogeneity of the degree
distribution is increased.  A pronounced minimum for the eigenratio $R$ at $\beta=1$
is also observed in various other models of complex networks \cite{MCJ1,MCJ2}.

\subsection{Mean Degree Approximation}
\label{s4a}

In what follows, we present an approximation for the
eigenratio $R$. Here we focus on the case of maximum
synchronizability ($\beta=1$).

Based on results of Ref.~\cite{CLV:2003} for random networks with
given expected degrees, we have
\begin{equation}
\max \{1-\lambda_2,\lambda_N-1\}= [1+o(1)]\frac{2}{\sqrt{k}}.
\label{eq24}
\end{equation}
Moreover, the semicircle law holds and the spectrum of matrix $H$ is symmetric
around $1$ for $k_{\min}\gg \sqrt{k}$ in the thermodynamic limit \cite{CLV:2003}.
These results are rigorous for ensembles of networks with a given expected
degree sequence and sufficiently large minimum degree $k_{\min}$, but our numerical
computation supports the hypothesis
that the approximate relations
\begin{equation}
\lambda_2 \approx 1 - \frac{2}{\sqrt{k}}, \;\;\; 
\lambda_N \approx 1 + \frac{2}{\sqrt{k}},
\label{eq25}
\end{equation}
hold under much milder conditions.  In particular, relations (\ref{eq25}) are
expected to hold true for any large, sufficiently random network with $k_{\min}\gg 1$.

Under the assumption that $1-\lambda_2\approx \lambda_N-1 \approx  2/\sqrt{k}$, the eigenratio
can be written as
\begin{equation}
R \approx \frac{1+2/\sqrt{k}}{1-2/\sqrt{k}}.
\label{eq27}
\end{equation}
Therefore, for $\beta=1$, the eigenratio $R$ is primarily determined by the mean
degree and does not depend on the number of oscillators and the details of the
degree distribution. This is a remarkable result because, regardless of the degree
distribution, the network at $\beta=1$ is just as synchronizable as a random
homogeneous network with the same mean degree, and random homogeneous networks
appear to be asymptotically optimal in the sense that $R$ approaches the absolute
lower bound in the thermodynamic limit for large enough $k$ \cite{Wu:2003}.

\begin{figure}[pt]
\begin{center}
\epsfig{figure=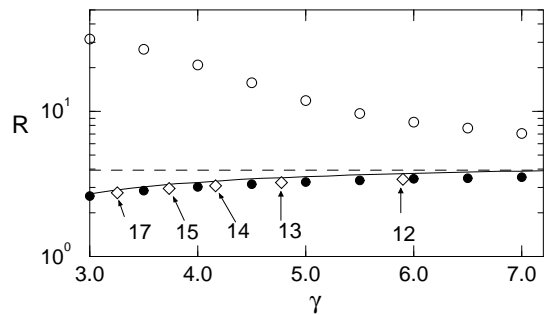,width=7.0cm}
\caption{
 Eigenratio $R$ as a function of the scaling exponent $\gamma$ for random
 SFNs with
 $\beta=1$ ($\bullet$) and $\beta=0\;$ ($\circ$). The other curves are
 the approximation of the eigenratio in Eq.~(\ref{eq27}) (solid line)
 and the eigenratio at $\gamma=\infty$ (dashed line).
 The ${\scriptstyle \lozenge}$ symbols correspond
 to random homogeneous networks with the same mean degree of the
 corresponding SFNs (the degrees are indicated in the figure).
 The other network parameters are the same as in Fig.~\ref{fig1}.
}
\label{fig2}
\end{center}
\end{figure}

We now test our predictions on random SFNs.  As shown in Fig.~\ref{fig2}, in
unweighted SFNs, the eigenratio $R$ increases with increasing heterogeneity of
the degree distribution (see also Ref.~\cite{NMLH:2003}).  But, as shown in the
same figure, the eigenratio does not increase with heterogeneity when the
coupling is weighted at $\beta=1$.  The difference is particularly large for
small scaling exponent $\gamma$, where the variance of the degree distribution
is large and the network is highly heterogeneous (note the logarithmic scale in
Fig.~\ref{fig2}).  For $\beta=1$, the eigenratio $R$ is well approximated by the
relation in Eq.~(\ref{eq27}) [Fig.~\ref{fig2}, solid line].  For $\beta=1$, the
eigenratio of the SFNs is also very well approximated by the eigenratio of
random homogeneous networks with the same number of links [Fig.~\ref{fig2},
${\scriptstyle \lozenge}$].  Therefore, for $\beta=1$, the variation of the
eigenratio $R$ with the heterogeneity of the degree distribution in SFNs is
mainly due to the variation of the mean degree of the networks, which increases
in random SFNs as the scaling exponent $\gamma$ is reduced.

\begin{figure}[pt]
\begin{center}
\epsfig{figure=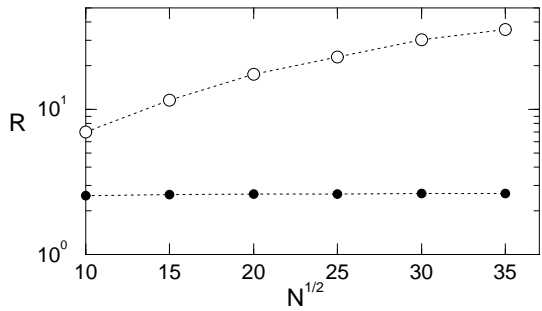,width=7.0cm}
\caption{Eigenratio $R$
 as a function of the number of oscillators for random SFNs with $\gamma=3$.
 Dotted lines are guides for the eyes.
 The legend and other parameters are the same as in Fig.~\ref{fig2}.
}
\label{fig2.5}
\end{center}
\end{figure}

In Fig.~\ref{fig2.5}, we show the eigenratio $R$ as a function of the system
size $N$.  In unweighted SFNs, the eigenratio increases strongly as the number
of oscillators is increased.  Therefore, it may be very difficult or even
impossible to synchronize large unweighted networks.  However, for $\beta=1$,
the eigenratio of large networks appears to be independent of the system size,
as shown in Fig.~\ref{fig2.5} for random SFNs.  Similar results are observed in
many other models of complex networks.  Altogether, these provide strong
evidence for relation (\ref{eq27}) and shows that synchronizability is
significantly enhanced for $\beta=1$ as compared to the case of unweighted
coupling ($\beta=0$).

\subsection{Coupling Cost}

We now turn to the problem of the cost involved in the
network of couplings.  The total cost $C$ is
naturally defined as the total input strength of the connections of all
nodes at the synchronization threshold: 
\begin{equation}
 C=\sigma_{\min}\sum_{i=1}^{N}k_i^{1-\beta},
\label{eq30}
\end{equation}
where $\sigma_{\min}=\alpha_1/\lambda_2$ is the minimum coupling strength
for the network to synchronize.  We recall that $\alpha_1$ is the point
where the master stability function first becomes negative.  
For $\beta=1$, we have $C=N\alpha_1/\lambda_2$.

\begin{figure}[pt]
\begin{center}
\epsfig{figure=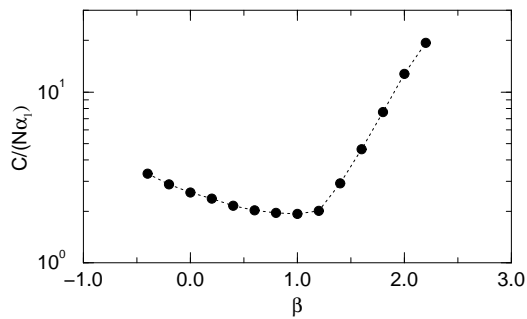,width=7.0cm}
\caption{Normalized cost $C/N\alpha_1$ as a function of $\beta$
for random SFNs with scaling exponent $\gamma=3$. The other parameters
are the same as in Fig.~\ref{fig1}.
}
\label{fig5}
\end{center}
\end{figure}

\begin{figure}[pt]
\begin{center}
\epsfig{figure=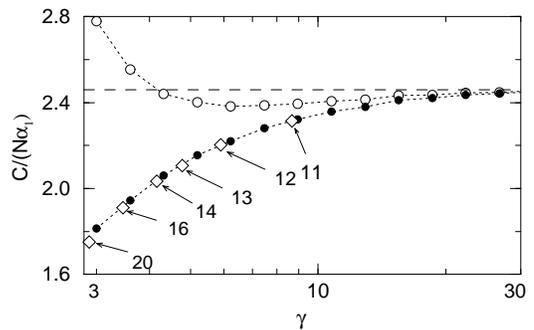,width=7.0cm}
\caption{Normalized cost $C/N\alpha_1$ as a function of the scaling exponent $\gamma$ for
random SFNs with $\beta=1$ ($\bullet$) and $\beta=0$ ($\circ$),
and for random homogeneous networks with the same mean degree (${\scriptstyle
\lozenge}$).  The  dashed line corresponds to $\gamma=\infty$.
The other parameters are the same as in Fig.~\ref{fig1}.
}
\label{fig4}
\end{center}
\end{figure}

As a function of $\beta$, the cost  has a  broad minimum at $\beta=1$, as
shown in Fig.~\ref{fig5} for random SFNs. 
This result is important because it shows that maximum
synchronizability and minimum cost occur exactly at the same point.
The cost for random SFNs at $\beta=1$ is very well approximated by
the cost for random homogeneous networks with the same mean degree
[Fig.~\ref{fig4}, ${\scriptstyle \lozenge}$], in agreement with our analysis in
Sec.~\ref{s4a} that,
at $\beta=1$, the eigenvalue $\lambda_2$ is fairly independent of the degree distribution. 
The cost at $\beta=1$ is significantly reduced as
compared to the case of unweighted coupling ($\beta=0$), as shown in Fig.~\ref{fig4}.
The difference becomes more pronounced when
the scaling exponent $\gamma$ is reduced and the degree distribution becomes
more heterogeneous.
Similar results are observed in other models of complex networks. 
Therefore, cost reduction is another important advantage of suitably weighted networks.

\section{Phase synchronization}
\label{s4}

The above analysis on identical oscillators can serve as a
good approximation even when the oscillators are not fully
identical. In some realistic situations~\cite{PRK:book}, however, the
interacting oscillators are significantly different, specially
in the oscillation frequencies~\cite{RPK}. Here we extend the enhancement
of synchronizability to networks of non-identical oscillators.

We consider phase synchronization in complex networks of limit cycle oscillators,
\begin{equation}
\dot{\bf x}_i = \omega_i{\bf F}({\bf x}_i)-\sigma\sum_{j=1}^{N}G_{ij} {\bf H}({\bf x}_j), \;\;\; i=1,\ldots ,N, 
\label{peq1}
\end{equation}
where $\dot{\bf x}= {\bf F}({\bf x})$ represents a Van der Pol oscillator, 
$\dot{u}=v$, $\dot{v}=0.3(1-u^2)v-u$,
and the couplings are through both variables $u$ and $v$
such that ${\bf H}({\bf x})={\bf x}$.
We assume that the frequencies $\omega_i$ follow a uniform distribution in an interval
$[1-\Delta \omega, 1+\Delta \omega]$. In our simulations we set $\Delta \omega=0.2$.

\begin{figure}[pt]
\begin{center}
\epsfig{figure=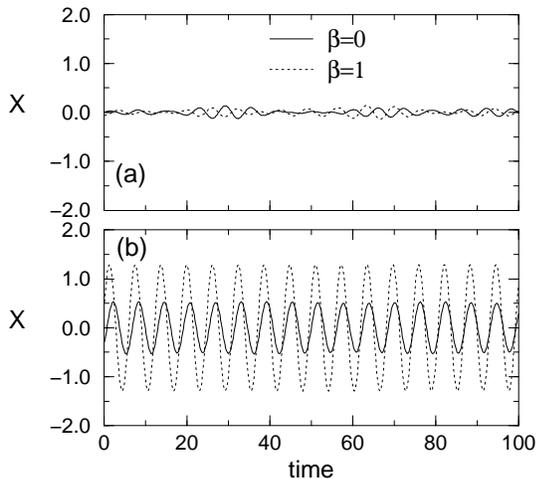,width=7.0cm}
\caption{Time series of the mean field $X$ in random SFNs
of non-identical Van der Pol oscillators with  $\gamma=3$.
The coupling strengths are (a) $\sigma^*=0.1$ and (b) $\sigma^*=0.2$.
The other parameters are the same as in Fig.~\ref{fig1}.
}
\label{fig61}
\end{center}
\end{figure}

Now we study the effects of 
the weighted coupling on phase synchronization. For an arbitrary $\beta$, 
on average each oscillator receives an effective  input of strength  
$\sigma^*=\sigma\sum_i^N k_i^{1-\beta}/N$. We can compare the synchronization 
behavior at different $\beta$ for the same effective strength $\sigma^*$.
In Fig.~\ref{fig61}, we show time series of the mean field $X=\sum_{i=1}^{N}x_i/N$
for unweighted ($\beta=0$) and weighted random SFNs ($\beta=1$). For small coupling
strength $\sigma^*$, neither of the networks display significant collective behavior
[Fig.~\ref{fig61}(a)]. As the coupling strength is increased, collective oscillations
emerge for both weighted and unweighted networks, but the oscillations are much
more pronounced for the networks with $\beta=1$ [Fig.~\ref{fig61}(b)].

\begin{figure}[pt]
\begin{center}
\epsfig{figure=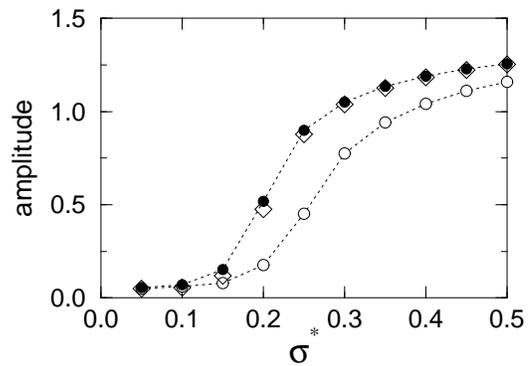,width=7.0cm}
\caption{Amplitude as a function of $\sigma^*$
for random SFNs of Van der Pol oscillators 
with $\beta=0$ ($\circ$) and $\beta=1$ ($\bullet$).
The ${\scriptstyle \lozenge}$ symbols correspond to 
random homogeneous networks with the same mean degree
of the random SFNs. The results are averaged over $20$ realizations.
The scaling exponent is $\gamma=3$ and the other parameters are
the same as in Fig.~\ref{fig1}.
}
\label{fig62}
\end{center}
\end{figure}

In Fig.~\ref{fig62}, we show the {\it amplitude} of the mean field $X$ as a function
of $\sigma^*$. Here the amplitude is
defined as the standard deviation of $X$ over time. The amplitude is approximately
zero for small coupling strength, increases sharply as $\sigma^*$ is increased
beyond a certain critical value, and saturates for large $\sigma^*$ (Fig.~\ref{fig62}). 
(This transition becomes sharper as the number $N$ of oscillators is increased).  
The overall behavior is similar for weighted and unweighted networks but, again,
the amplitude is significantly larger for networks with  $\beta=1$ at the same value
of $\sigma^*$. 
Moreover, the amplitude of random SFNs at
$\beta=1$ is well approximated by the amplitude of random homogeneous networks
with the same mean degree (Fig.~\ref{fig62}), which are networks that exhibit
good phase synchronization properties. 

All these provide strong evidence that our weighted coupling scheme
strongly enhances the synchronizability of networks of non-identical oscillators.
This is expected to be relevant for realistic networks.

\section{What are the most synchronizable networks?}
\label{s5}

Within the weighted coupling scheme of Eq.~(\ref{eq2}),
the networks that exhibit maximum synchronizability and
minimum cost are those with $\beta =1$ (Sec.~\ref{s3}).
Physically, this seems to be related to the fact that
the synchronization of the network is determined by the
input signal at each node (as opposed to the output signal)
and that the sum $\sum_j A_{ij}=k_i^{1-\beta}$
of the strengths of all the in-links of a node becomes
uniformly equal to $1$ for all the nodes exactly at $\beta =1$.
This leads to the natural question of whether there
is any other distribution of weights that can further
improve the synchronizability and cost of the network
for a given topology of connections. As we show,
the answer is positive for finite size networks, but
seems to be negative in the thermodynamic limit.

In order to address this question we introduce the following
weighted coupling scheme.
If there is a link between a node $i$ with degree $k_i$
and a node $j$ with degree $k_j$, the strength $A_{ij} = w_{ij}$
of link $j\rightarrow i$ is taken to be
\begin{equation}
w_{ij}= (k_ik_j)^{\theta}/N_r
\label{neq1}
\end{equation}
where $\theta$ is a tunable parameter and
$N_r=\sum_{j\sim i} (k_ik_j)^{\theta}$ is a normalization factor that
ensures that $\sum_j A_{ij}=1$ (the sum $\sum_{j\sim i}$ is over all the neighbors
of node $i$). 
For $\theta =0$ we recover the model in Eq.~(\ref{eq2}) for $\beta=1$.

\begin{figure}[pt]
\begin{center}
\epsfig{figure=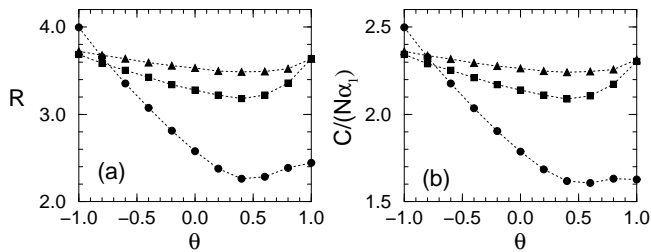,width=8.5cm}
\caption{(a) Eigenratio $R$ and (b) normalized cost $C/(N\alpha_1)$
as functions of $\theta$ for random SFNs with 
with $\gamma=3$ ($\bullet$), $\gamma=5$ (${\scriptscriptstyle \blacksquare}$),
and $\gamma=7$ (${\scriptstyle \blacktriangle}$). 
The other parameters are the same as in Fig.~\ref{fig1}.
}
\label{fig7}
\end{center}
\end{figure}

This new weighted model
is partially motivated by observations in real networks, including scientific
collaboration networks~\cite{BBPV:2004}, 
airport networks~\cite{BBPV:2004,MAB:2004},
and metabolic networks~\cite{MAB:2004}, which have
been found to exhibit a strong correlation between the
weight $w_{ij}$ and product of the corresponding degrees as  
$\langle w_{ij}\rangle \sim (k_ik_j)^\theta$. A similar model
has been considered in the study of coherence in networks of chaotic maps
\cite{lgh}.

We now study the weighted model (\ref{neq1}) numerically
in the context of complete synchronization of identical oscillators.
In Fig.~\ref{fig7}, we show the eigenratio $R$ and the normalized cost
$C/(N\alpha_1)$ as functions of $\theta$ for different values of the
scaling exponent. In each case, both the eigenratio and the cost reach a minimum
for some $\theta^*(N,\gamma)>0$, which is approximately the same for $R$ and $C$ and
indicates that these networks synchronize better than those of the weighted
model~(\ref{eq2}) for $\beta=1$ ($\theta=0$).
However, as compared to the difference between unweighted networks $(\beta=0)$
and networks weighted for $\beta=1$ in model~(\ref{eq2}), the further reduction
of $R$ and $C$ within model~(\ref{neq1}) is very small.

\begin{figure}[pt]
\begin{center}
\epsfig{figure=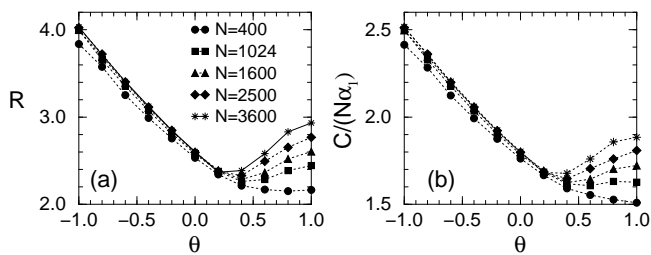,width=8.5cm}
\caption{The same as in Fig.~\ref{fig7} for random SFNs with
$\gamma=3$ and various choices for the number $N$ of oscillators,
as indicated in the figure.
}
\label{fig8}
\end{center}
\end{figure}

Indeed, $R(\beta=0)-R(\beta=1)$ and $C(\beta=0)-C(\beta=1)$ [Fig.~\ref{fig2}]
are orders of magnitude larger than
$R(\theta=0)-R(\theta=\theta^*)$ and $C(\theta=0)-C(\theta=\theta^*)$ [Fig.~\ref{fig7}],
even when the networks are strongly heterogeneous (i.e., $\gamma$ is small).
Moreover, the difference between the synchronizability at $\theta=0$ and $\theta=\theta^*$
reduces when the number $N$ of oscillators is increased,
as shown in Fig.~\ref{fig8}. Since $\theta^*$ approaches $\theta=0$ for increasing $N$,
our numerics are consistent with the hypothesis that the minimum of $R$ and $C$ will
be at $\theta=0$ for $N\rightarrow\infty$. That is, the weighted model~(\ref{eq2})
for $\beta=1$ may be optimal in the thermodynamic limit. The verification of this
hypothesis is currently an interesting open problem.\\

\section{Conclusions}
\label{s6}

Complex networks with strong heterogeneity in the distributions of either degrees
or weights are poorly synchronizable.  This suppression of synchronizability is
due to the different intensities of the input signals received by different nodes
in the network.  We have shown that the synchronizability is significantly improved
when the distribution of weights is constrained by the distribution of degrees
in such a way that all the nodes receive the same intensity of input signal.
In this case, the distributions of degrees and weights as well as the distribution
of the total strength of out-links per node, $\sum_i A_{ij}$, may be highly
heterogenous, but the distribution of the total strength
of in-links, $\sum_j A_{ij}$, is the same for all the nodes in the network.
In particular, our analysis shows that properly weighted complex networks are much
more synchronizable than the corresponding unweighted networks.

\acknowledgements

This work was partially supported by SFB 555 and the VW Foundation.


\begin{references}

\bibitem{rev}
S. H. Strogatz, Nature (London) {\bf 410}, 268 (2001);
R. Albert and A.-L. Barab\'{a}si, Rev. Mod. Phys. {\bf 74}, 47 (2002);
S. N. Dorogovtsev and J. F. F. Mendes, Adv. Phys. {\bf 51}, 1079 (2002);
M. E. J. Newman, SIAM Rev. {\bf 45}, 167 (2003). 

\bibitem{watts:book}
D. J. Watts, {\it Small Worlds} (Princeton University Press, Princeton, 1999).

\bibitem{sync}
L. F. Lago-Fern\'andez, R. Huerta, F. Corbacho, and J. A. Sig\"uenza, Phys. Rev. Lett. {\bf 84}, 2758 (2000);
P. M. Gade and C. K. Hu, Phys. Rev. E {\bf 62}, 6409 (2000);
L. G. Morelli and D. H. Zanette, Phys. Rev. E {\bf 63}, 036204 (2001);
K. Sun and Q. Ouyang,  Phys. Rev. E  {\bf 64}, 026111 (2001);
J. Jost and M. P. Joy,  Phys. Rev. E {\bf 65}, 016201 (2002);
X. F. Wang, Int. J. Bifurcation Chaos Appl. Sci. Eng. {\bf 12}, 885 (2002);
H. Hong, M. Y. Choi, and B. J. Kim, Phys. Rev. E {\bf 65}, 026139 (2002);
O. Kwon and H. T. Moon, Phys. Lett. A {\bf 298}, 319 (2002);
M. Timme, F. Wolf, and T. Geisel, Phys. Rev. Lett. {\bf 89}, 258701 (2002);
G. W. Wei, M. Zhan, and C. H. Lai, Phys. Rev. Lett. {\bf 89}, 284103 (2002);
F. Oi, Z. Hou, and H. Xin, Phys. Rev. Lett. {\bf 91}, 064102 (2003);
J. Ito and K. Kaneko, Phys. Rev. E {\bf 67}, 046226 (2003);
A. M. Batista, S. E. D. Pinto, R. L. Viana, and S. R. Lopes, Physica A {\bf 322}, 118 (2003);
F. M. Atay, J. Jost, and A. Wende, Phys. Rev. Lett. {\bf 92}, 144101 (2004);
Y. Moreno and A. F. Pacheco,  Europhys. Lett. {\bf 68}, 603 (2004).

\bibitem{lgh}
P. G. Lind, J. A. C. Gallas, and H. J. Herrmann, Phys. Rev. E {\bf 70}, 056207 (2004).

\bibitem{GDLP:2000}
X. Guardiola, A. Diaz-Guilera , M. Llas, and C. J. Perez, Phys. Rev. E {\bf 62}, 5565 (2000).
 
\bibitem{BP:2002}
M. Barahona and L. M. Pecora, Phys. Rev. Lett. {\bf 89}, 054101 (2002).

\bibitem{NMLH:2003}
T. Nishikawa, A. E. Motter, Y.-C. Lai, and F. C. Hoppensteadt, Phys. Rev. Lett. {\bf 91}, 014101 (2003). 

\bibitem{Wu:2003}
C. W. Wu, Phys. Lett. A {\bf 319}, 495 (2003).

\bibitem{heter}
M. Denker, M. Timme, M. Diesmann, F. Wolf, and T. Geisel, Phys. Rev. Lett. {\bf 92}, 074103 (2004).

\bibitem{ROH:2004}
J. G. Restrepo, E. Ott, and B. R. Hunt, Phys. Rev. E {\bf 69}, 066215 (2004).

\bibitem{sw}
D. J. Watts and S. H. Strogatz, Nature (London) {\bf 393}, 440 (1998).

\bibitem{sf}
A.-L. Barab\'{a}si and R. Albert, Science {\bf 286}, 509 (1999).

\bibitem{com1}
A different behavior has been observed in networks of pulse oscillators \cite{GDLP:2000}.

\bibitem{spath}
F. Chung and L. Lu, Proc. Natl. Acad. Sci. U.S.A. {\bf 99}, 15879 (2002);
B. Bollob\'{a}s and O. Riordan, in Handbook of Graphs and Networks,
   edited by S. Bornholdt and H. G. Schuster (Wiley-VCH, Berlin, 2002);
R. Cohen and S.  Havlin, Phys. Rev. Lett. {\bf 90}, 058701 (2003);
S. N. Dorogovtsev, J. F. F. Mendes, and A. N. Samukhin, Nucl. Phys. B {\bf 653}, 307 (2003).

\bibitem{BBPV:2004}
A. Barrat, M. Barth\'elemy, R. Pastor-Satorras, and A. Vespignani, Proc. Natl. Acad. Sci. U.S.A. {\bf 101}, 3747 (2004).

\bibitem{MAB:2004} 
P. J. Macdonald, E. Almaas, A.-L. Barab\'{a}si, e-print cond-mat/0405688.

\bibitem{w}
S. H. Yook, H. Jeong, A.-L. Barab\'{a}si, and Y. Tu, Phys. Rev. Lett. {\bf 86}, 5835 (2001).
M. E. J. Newman, Phys. Rev. E {\bf 64}, 016132 (2001);
H. J. Kim, I. M. Kim,  Y. Lee, and B. Kahng, J. Korean Phys. Soc. {\bf 40}, 1105 (2002);
V. Latora and M. Marchiori, Eur. Phys. Journ. B {\bf 32}, 249 (2003);
C. Aguirre, R. Huerta, F. Corbacho, and P. Pascual, Lecture Notes in Computer Science, Vol. 2415 (2002), p. 27;
L. A. Braunstein, S. V. Buldyrev, R. Cohen, S. Havlin, and H. E. Stanley, Phys. Rev. Lett. {\bf 91}, 168701 (2003);
C. Li and G. Chen, e-print cond-mat/0311333;
G. Caldarelli, F. Coccetti, and P. De Los Rios, e-print cond-mat/0312236.


\bibitem{brain}
D. J. Felleman and D. C. Van Essen, Cerebral Cortex {\bf 1}, 1 (1991); 
J. W. Scannell,  G. A. P. C. Burns, C. C. Hilgetag, M. A. O'eil, and M. P. Yong,, Cerebral Cortex {\bf 9}, 277 (1999).


\bibitem{city} 
B. T. Grenfell, O. N. Bjornstad, and J. Kappey, Nature (London) {\bf 414}, 716 (2001).

\bibitem{MCJ1}
A. E. Motter, C. S. Zhou, and J. Kurths, Europhys. Lett. {\bf 69}, 334 (2005).

\bibitem{MCJ2}
A. E. Motter, C. S. Zhou, and J. Kurths, Phys. Rev. E {\bf 71}, 016116 (2005). 

\bibitem{chung:book}
F. R. K. Chung, {\it Spectral Graph Theory} (AMS, Providence, 1994).

\bibitem{other_ref}
M. Faloutsos, P. Faloutsos, and C. Faloutsos, Comput. Commun. Rev. {\bf 29}, 251 (1999);
R. Monasson, Eur. Phys. J. B, {\bf 12} 555 (1999);
I. J. Farkas,  I. Der\'{e}nyi, A.-L.  Barab\'{a}si, and T. Vicsek, Phys. Rev. E {\bf 64}, 026704 (2001);
K. I. Goh, B. Kahng, and D. Kim, Phys. Rev. E {\bf 64}, 051903 (2001);
Z. Burda, J. D. Correia, and A. Krzywicki, Phys. Rev. E {\bf 64}, 046118 (2001);
M. Mihail, C. Gkantsidis, and E. Zegura, {\it Spectral analysis of Internet topologies}, in Proc. Infocom. IEEE (2003).

\bibitem{CLV:2003}
F. Chung, L. Lu, and V. Vu, Proc. Natl. Acad. Sci. U.S.A. {\bf 100}, 6313 (2003).

\bibitem{DGMS:2003}
S. N. Dorogovtsev, A. V. Goltsev, J. F. F. Mendes, and A. N. Samukhin, Phys. Rev. E {\bf 68}, 046109 (2003).

\bibitem{msf}
L. M. Pecora and T. L. Carroll, Phys. Rev. Lett. {\bf 80}, 2109 (1998);
K. S. Fink, G. Johnson, T. Carroll, D. Mar, and L. Pecora, Phys. Rev. E {\bf 61}, 5080 (2000).

\bibitem{NSW:2001}
M. E. J. Newman, S. H. Strogatz, and D. J. Watts, Phys. Rev. E {\bf 64}, 026118 (2001).

\bibitem{PRK:book}
A. S. Pikovsky, M. G. Rosenblum, and J. Kurths, {\it Synchronization: A universal concept in nonlinear
sciences} (University Press, Cambridge, 2001).

\bibitem{RPK}
M. G. Rosenblum, A. S. Pikovsky, and J. Kurths, Phys. Rev. Lett. {\bf 76}, 1804 (1996). 

\end{references}
\end{document}